\documentclass{ieeeaccess}
\usepackage{cite}
\usepackage{amsmath,amssymb,amsfonts}
\usepackage{algorithmic}
\usepackage{graphicx}
\usepackage{textcomp}
\usepackage{wrapfig}

\def\BibTeX{{\rm B\kern-.05em{\sc i\kern-.025em b}\kern-.08em
    T\kern-.1667em\lower.7ex\hbox{E}\kern-.125emX}}
\begin{document}
\history{Date of publication xxxx 10, 2021, date of current version xxxx 10, 2021.}
\doi{02.2021/ArXiv.org}


\title{Poisoning Attacks and Defenses on Artificial Intelligence: A Survey}

\author{{Miguel A. Ramirez}\authorrefmark{1 $\dagger$}, {Song-Kyoo Kim}\authorrefmark{1,2 $\dagger$}, {Hussam Al Hamadi}\authorrefmark{1}, {Ernesto Damiani}\authorrefmark{1}, Young-Ji Byon\authorrefmark{3}, Tae-Yeon Kim\authorrefmark{3}, Chung-Suk Cho\authorrefmark{3} and {Chan Yeob Yeun}\authorrefmark{1}}

\address[1]{Center for Cyber-Physical Systems, Khalifa University of Science and Technology, Abu Dhabi, UAE.}
\address[2]{School of Applied Sciences, Macao Polytechnic Institute, R. de Luis Gonzaga Gomes, Macao, SAR.}
\address[3]{Department of Civil Infrastructure and Environmental Engineering, Khalifa University of Science and Technology, Abu Dhabi, UAE.}
\address[$\dagger$]{The first two authors have equal contribution.}


\markboth
{MA Ramirez, SK Kim, \headeretal: PA and Defenses, ArXiv.org, 2022.}
{MA Ramirez, SK Kim, \headeretal: PA and Defenses, ArXiv.org, 2022.}


\begin{abstract}
Machine learning models have been widely adopted in several fields. However, most recent studies have shown several vulnerabilities from attacks with a potential to jeopardize the integrity of the model, presenting a new window of research opportunity in terms of cyber-security. This survey is conducted with a main intention of highlighting the most relevant information related to security vulnerabilities in the context of machine learning (ML) classifiers; more specifically, directed towards training procedures against data poisoning attacks, representing a type of attack that consists of tampering the data samples fed to the model during the training phase, leading to a degradation in the model’s overall accuracy during the inference phase. This work compiles the most relevant insights and findings found in the latest existing literatures addressing this type of attacks. Moreover, this paper also covers several defense techniques that promise feasible detection and mitigation mechanisms, capable of conferring a certain level of robustness to a target model against an attacker. A thorough assessment is performed on the reviewed works, comparing the effects of data poisoning on a wide range of ML models in real-world conditions, performing quantitative and qualitative analyses. This paper analyzes the main characteristics for each approach including performance success metrics, required hyperparameters, and deployment complexity. Moreover, this paper emphasizes the underlying assumptions and limitations considered by both attackers and defenders along with their intrinsic properties such as: availability, reliability, privacy, accountability, interpretability, etc. Finally, this paper concludes by making references of some of main existing research trends that provide pathways towards future research directions in the field of cyber-security.

\end{abstract}


\titlepgskip=-15pt

\maketitle

\section{Introduction}
The reliability associated with modern artificial intelligence (AI) models plays an important role in a wide range of applications \cite{Barreno, A01} including Internet-of-Things \cite{Olufowobi}. Consequently, over the last couple of years, these machine learning (ML) models demand adopting additional techniques in order to address security related issues since newly emerging vulnerabilities are being discovered and capable of posing a threat to the integrity of the ML model which is the target of an attacker \cite{R04}. An attacker could exploit such vulnerabilities causing a negative impact on the performance of the ML model. It has been proven plausible to maliciously compromise a training dataset in order to affect the decision-making process of the model which can cause malfunctions during testing (i.e. inference) phase \cite{R04}. 
The need of public and available data is continuously on demand by most ML models. A clear example can be seen in smart city systems wherein large amounts of data are gathered by numerous sensors, such as smartphones. Then it can be foreseen that the consequences of an attack targeting smart city systems could be devastating and such an event is prone to occur due to the system being heavily dependent on public data. The main objective of this survey is directed towards gathering some of the most representative attack and defense approaches from the perspective of data poisoning \cite{Paudice_1}. Data poisoning (DP) attacks aim to compromise the integrity of a target model by performing alterations to the required dataset used by the model during the training phase. This causes the model to misclassify samples during the testing phase, resulting in a significant reduction in the overall accuracy. Due to the above mentioned considerations, there is an urge to develop more advanced defense mechanisms, aiming to enhance the robustness of the model against potential DP attacks occurring while training, in order to mitigate the effects of the data poisoning. It is expected that a profound analysis over the latest advances in defense schemes against poisoning attacks could serve as a guideline for developing a novel approach that attains a certain level of immunization against DP on smart devices feeding data to smart city systems. The main contributions of this survey compared to other existing review/survey papers are listed as follows:

\ 

\begin{itemize}
\item[(1)] This paper reviews related works of machine learning security mechanisms, focusing primarily on potential threats involving poisoning attacks occurring during the training phase, particularly towards manipulation of mislabeled training data.

\item[(2)] Several threats and attack strategies for machine learning (ML) models are presented and analyzed, describing capabilities of the attacks and the list of assumptions involved in each attack of interest. Furthermore, attacks on the ML models are classified into 2 categories: Attacks on Non-Neural Networks (NN) and Attacks on Neural Networks.

\item[(3)] Various defense techniques are examined, highlighting their potential benefits as well as their disadvantages or challenges entailing details of their deployments.

\item[(4)] This paper suggests future research directions for the field of machine learning security with discussions of their implications.
\end{itemize}

\ 

This survey paper is organized as follows: the related background knowledge is explained on Section II. The Various types of data poisoning attacks and their defense mechanisms are presented on Section III and IV. Section V discusses and suggests research directions in the recent studies which have been reviewed by our survey. Overall lessons and the insights of the survey are covered in Section VI. Basic statistics of the survey is also included in Section VI. Section VII concludes this paper. 

\section{Knowledge Background}
In this section, an overview of the properties related to attacks and defenses is presented. The fundamentals of relevant topics involving the security issue in machine learning models are discussed, mainly from the perspective of assumptions about attackers and different types of attacks throughout the machine learning lifecycle.

\subsection{Manipulations}
Training data manipulation \cite{A02} is one of different types of DP attacks by corrupting (or poisoning) the training data during a training phase with an aim of utterly jeopardizing the integrity of the ML classifier making it to become an ineffective classifier.  Examples of techniques often used by the attackers are the modification of data labels, injection of malicious samples and manipulation of the training data. As a result, the overall damage to the target ML model can only be prevailed at a later inference phase, with the accuracy of the model being drastically reduced. This effect is commonly referred to as an accuracy degradation. 

\ 

An input manipulation refers to triggering a machine learning system to malfunction by altering the input that is fed into the system \cite{A03}. It would be in a form of an altered image adding noises or another input that causes the classifier to perform towards making wrong predictions. Adversarial attacks \cite{Miller_Survey, X_Ma, Y_Ma, R02, R12} take place during the inference phase, when the previously trained ML model is considered to be reliable and is assumed to perform with high accuracies \cite{R09}. Depending on the goal of the attacker, an adversarial attack can fall in one of two categories. The first category is referring to a targeted attack when the input in a form of crafted adversarial examples leads to the target model to misclassify the samples into a specific class defined by the attacker \cite{D_Meng, R01}. In contrast, in a Non-targeted attack the crafted adversarial examples aim to cause the target model to misclassify. Nonetheless, there is no need nor interest from the attacker to misclassify into a particular class apart from the correct one. Evasion attacks are also another kind of input manipulation and are different from adversarial attacks in a sense that evasion attacks do not require any knowledge over the training data \cite{R05}.

\subsection{Assumptions of attack and defense}
Security threats to machine learning models are generally divided into data poisoning (DP) attacks and adversarial attacks, the former is applied during a training phase and the latter is applied during a testing phase, this difference is shown in Figure \ref{DP}. For the purposes of this paper, data poisoning attacks will remain as the main topic of interest. 

\Figure[!h][width=0.4\textwidth]{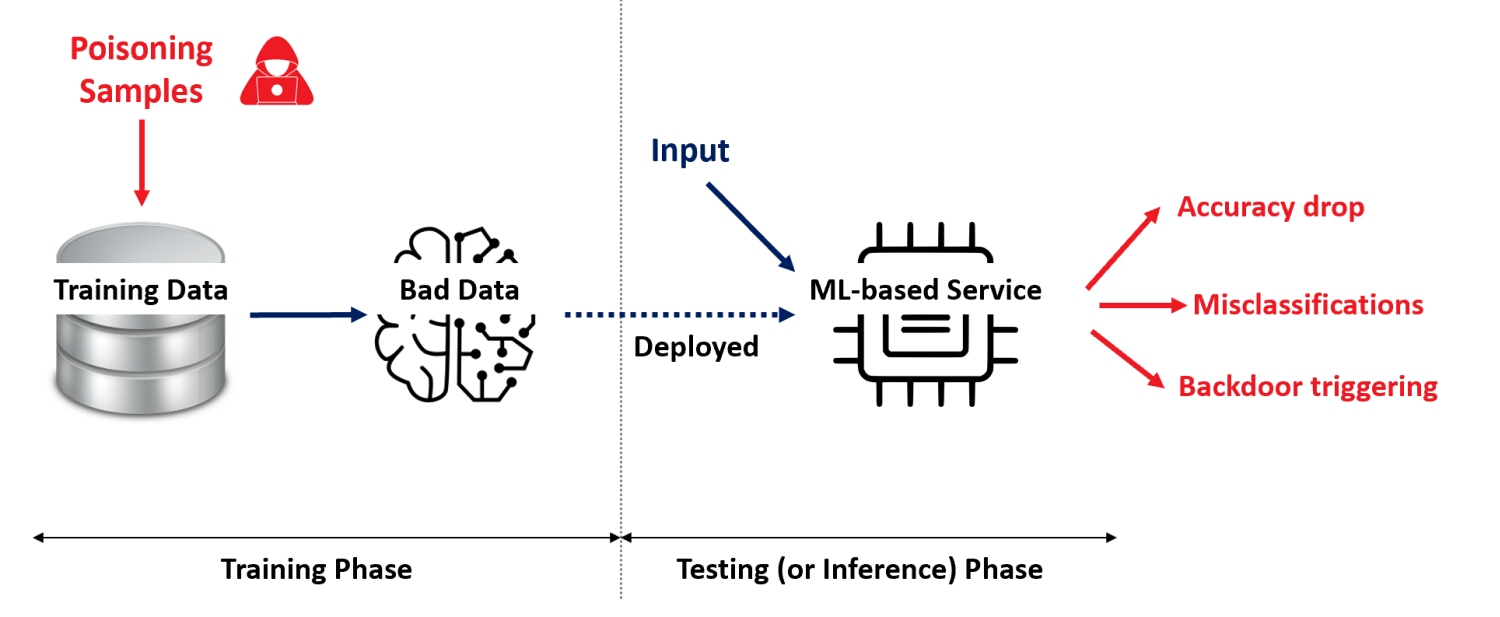}{Data poisoning attacks during training phase affecting testing phase \cite{Liu_Ximeng}. \label{DP}}

One of the most common schemes of the attacker is injecting malicious samples into the target model’s training set, corrupting either the feature values or labels of the training samples, and affecting the ML model boundaries by causing significant deviations to a point where the model’s reliability is completely devastated. As a result, this would leave the model susceptible to make wrong predictions. In summary, the main goal of a such attacker is to disrupt the training process aiming to significantly reduce the performance of the target model, causing a degradation in accuracy; and increasing misclassification rates of the samples during the testing phase.

\ 

Assumptions of attackers refer to the prior knowledge (implicit or explicit) about the target model of interest of the attacker, entailing the resources available to the attacker. When conducting experiments, the devised attack is meant to be evaluated against a defense, both attack and defense states-related assumptions must be declared to determine the conditions that guarantee either scheme’s efficiencies (e.g. attaining the defense to defeat an attack, or vice versa). However, various DP attacks have been shown to be successful in spite of having very little knowledge of the target model. An example of this is described in the research \cite{Lowd}, which directs a DP attack scheme to naive Bayes email-spam filters by simply sending ‘hamlike’ emails using a black-listed IP address as the sender, followed by being threatened and labeled as a spam, nonetheless these corrupt data will be inevitably used by the spam filter for further training. 

\ 

The assessment of the influence of the attacker over the training data is commonly defined as an attacker’s capability. Primary interest of the attacker is to alter either the feature values or labels as part of the training set. Nevertheless, the attacker is usually restricted to poison a limited number of samples, typically corresponding to a ratio of less than 30\% of the total data samples. More optimized poisoning algorithms have been in continuous development during the last decade, aiming to maximize the accuracy degradation and minimize the number of poisoning samples needed to perform the attack.

\subsection{Security and Reliability Requirements of AI models}
In this section, we define some of the most relevant qualitative security properties present in ML models \cite{A04}, each of these properties can be associated to a security threat and can be quantified by specific metrics \cite{R14}.

\

\begin{itemize}
\item \textbf{Integrity} is defined as an ability of a model to function according to specified norms in an understandable and predictable manner. Thereby, an attacker could tamper with the model’s parameters in the training phase and in turn affect the overall integrity of the model. This can be seen in the number of inputs representing label-flipping, which is the ratio of the models parameters impacted by attacking with respect to the target model during the training or testing phase.
\item \textbf{Availability} is associated to the ability of an ML model to perform as expected when facing radical perturbations with the potential of causing a considerable impact on the input data distribution due to the arising of unexpected conditions. Analyzing the ML model’s decision boundary represents a clear indicator of availability, which is reflected in the accuracy metric as well. For instance, the effect of a DP attack is magnified following a function of the ratio of the injected poisoning data samples in the training set, which leads to the utter breakdown of the decision boundary.
\item \textbf{Robustness} is defined as an ability of the model carrying on procedures in a desirable way in spite of having perturbations in the input distributions. Such perturbations could be deliberately crafted by an attacker, then performing the training of a model with poisoning samples substantially tampers with the model’s robustness. The distance from the last best version of the model can be used as a metric, as well as for ROC (Receiver Operating Characteristics) curve and AUC (Area Under the Curve).
\end{itemize}

\ 

For the purposes of this paper, the integrity is considered as the most important property as it is discussed in nearly all existing DP attack review literatures.

\subsection{Metrics of interest}
The success of DP attacks is measured based on the amount of degradation shown by the target model performance during the testing phase. This can be further verified after computing the decision matrix, observing the overall misclassification rates in each class displaying: true positive, false positive, true negative and false negative. Moreover, the effectiveness of the attack is shown in the form of a significant drop in the overall accuracy, this is referred as an accuracy degradation. The employment of additional metrics besides measuring the accuracy have been proposed in this paper to reflect and analyze in further detail about the overall performance of the target model and make comparisons to other performances of models. Various metrics for artificial intelligence have been proposed by multiple standard bodies including the International Organization for Standardization (ISO) \cite{A21} and the National Institute of Standards and Technology (NIST) \cite{A22}. The metrics for AI typically include the accuracy, the precision, the recall, the ROC and its area\footnote[1]{AROC: Area under Receiver Operating Characteristic} \cite{Carlini}.

\ 

Other approaches such as the one by Biggio and Roli \cite{B_Biggio_1} introduces security evaluation curves as a way to characterize the performance of a ML model against an intended attack considering various levels of knowledge from the attackers. Thus this approach accomplishes a comprehensive evaluation of the overall security of the model; and by doing so, enables another means to compare assorted defense techniques.

\subsection{Adversarial Capabilities}

In the testing phase, the attacker naturally will aim to attain further knowledge over the target model in order to increase the effectiveness of adversarial attacks, the attacker typically focuses on any of the following five factors: Feature space, classifier type (e.g. DNN or SVM), classifier learning algorithm, classifier learning hyperparameters, and training dataset. 

\ 

\begin{itemize}
\item A \textbf{white-box} assumption is commonly defined as a scenario in which the attacker does have complete knowledge over all the five elements already described previously, as well as any defense mechanism already set on top of the model \cite{Nasr,A05, A06}.

\item A \textbf{black-box} assumption is the opposite to white-box assumption, when no knowledge of the target model, albeit query it can be plausible. Nonetheless, it is important to remark that, just having access to the training data grants the upper hand to the attacker over any defender, representing this training data the unadulterated or ‘clean’ dataset, in question \cite{A07, A08, A09, Salem, R07}. 

\item A \textbf{gray-box} assumption is often referred as a middle ground between white-box and black-box scenarios, where the prior knowledge on the attacker’s side can include the feature space, the target classifier; this includes the model architecture, model parameters and the training dataset; however, the defense mechanism on top is unknown to the attacker. The gray-box setting usually is used to evaluate the defense against the adversarial attack \cite{Liu_Ximeng}.
\end{itemize}

\section{Poisoning Attacks}
In the following paragraphs several examples of data poisoning attacks on different types of models are discussed. For the purposes of the survey, the center of focus is directed towards data poisoning attacks performed during the training phase. The effects of every poisoning technique is then analyzed for classifier models only. Albeit there are several prior works that address poisoning techniques on regression models, these will not be considered as covered by the scope of this survey.

\subsection{Label flipping attacks}
The most common way to generate this kind of poisoning is by maliciously tampering the labels in the data \cite{X_Liu}, this can be easily achieved by just flipping labels, thus generating mislabeled data as shown in Figure \ref{LF}. Label flipping can be performed either randomly or specifically depending on the aims of the attacker; the former aims to reduce the overall accuracy of all classes, the later does not aim to perform significant accuracy reduction, rather it is focus on the misclassification of a determined class in particular. Paudice et al. \cite{Paudice_2} proposes an optimal label flipping poisoning attacks compromising machine learning classifiers. Label flipping actions are performed following an optimization formulation focused on maximizing the loss function of the target model. This approach is considered computationally intractable due to the inclusion of heuristic functions enabling the label flipping attacks to downscale the computational cost.

\Figure[!h][width=0.4\textwidth]{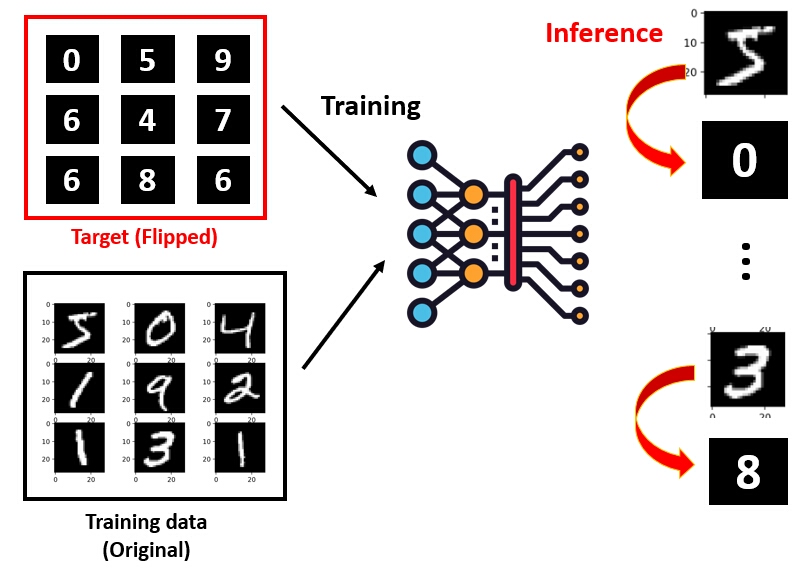}{Misclassification error caused by label-flipping \cite{X_Liu}.\label{LF}}

The applications of this approach limits itself to binary classification problems and the assumptions of the attack involves complete knowledge over the learning algorithm, loss function, training data and also the set of features used by the ML classifier, turning it basically into an attack on a White-box model. Albeit the list of assumptions appeal to unrealistic scenarios, the analysis emphasizes on worst case scenarios. The effectiveness of the propose method is demonstrated in three datasets from UCI repository: MNIST, Spambase and BreastCancer; succeeding in increasing the classification error by a factor of  6.0, 4.5 and 2.8, respectively \cite{Paudice_2}.

\ 

Xiao et al. \cite{H_Xiao_2} reports a successful attack on a SVM model after performing label flipping using an optimized framework capable of procure the label flips which maximizes potential classification errors, causing a significant reduction in the overall accuracy of the classifier. As a potential drawback, this technique naturally implies a high computational overhead as a main requirement.

\subsection{Attacks on Support Vector Machines (SVM)}
Support Vector Machines (SVMs) could be targeted for the poisoning attacks as shown in Figure \ref{SVM} \cite{D_Miller}. These attacks harness prior knowledge not only over the training data, but also onto the validation data and the hyperparameters of the SVM learning algorithm. Then the poisoning integrates an optimization method that maximizes the objective function based on the classifier error rate obtained over the validation data as a function of the location of the poisoning sample, this includes the class label as well. The optimization considers both the support and non-support vectors subsets are considered unaffected by the insertion of the poisoning samples during training.

\Figure[!h][width=0.3\textwidth]{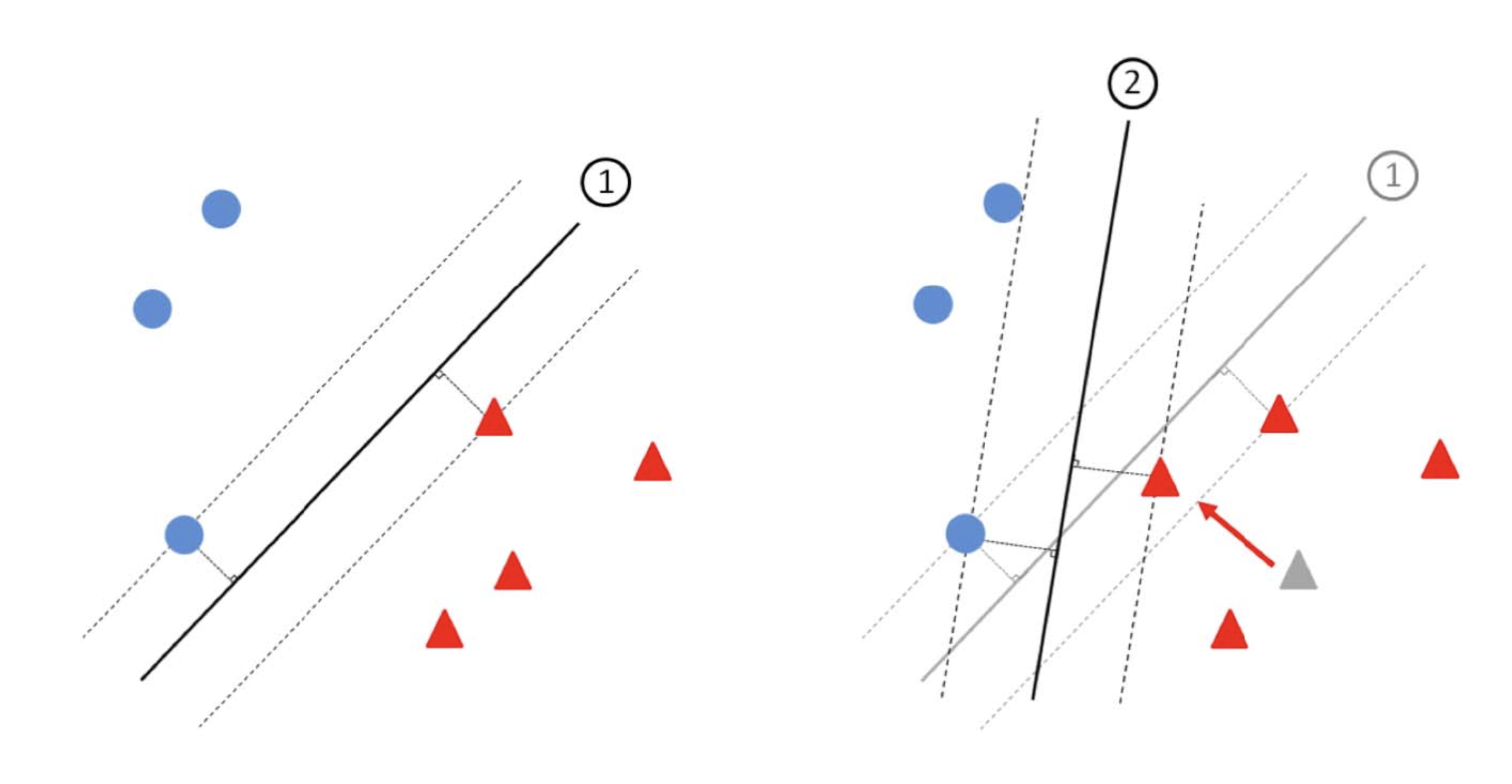}{(Left) SVM classifier decision boundary for two classes (Right) Impact on decision caused by the re-location of one single data point \cite{D_Miller}. \label{SVM}}

Biggio et at. \cite{B_Biggio_2} showcase a poisoning attack on SVM that requires as little as an insertion of one single poisoning sample to cause a considerable amount of accuracy degradation, demonstrating the high vulnerability of SVM models against DP attacks. After conducting an extensive evaluation of the target models performing over MNIST dataset, the reported error increased up to a range between 15\% and 20\% \cite{B_Biggio_2}. Poisoning attacks against SVM classifiers to increase the testing errors of the classifier, crafted training data is fed to the model has been also proposed \cite{B_Biggio_3}. Then based on the SVM model’s optimal solution, a gradient ascent strategy is deployed to construct the poisoning data. In addition, this method enables optimization formulation and allows itself to be kernelized. Nonetheless, such poisoning strategy requires full knowledge of both the algorithm of interest to the attacker and its training data. 

\subsection{Attacks on clustering algorithms}
Biggio et al. \cite{B_Biggio_5} performs a poisoning attack attempting against the clustering process with a reduced number of poisonous samples, assessing the effectiveness of the attack by performing evaluations of the target model on handwritten digits and malware samples, generating poisonous data samples by relying on a behavioral approach on malware clustering. The algorithms itself computes the existing distance between two clusters and deposits the poisoning samples right between the both of them, which in question creates conflicts with their decision boundaries, almost managing to merge both into a single cluster, as seen in \cite{B_Biggio_6}. As a result, the approach in both works generalize well among clustering models; nonetheless the approach is only valid in white-box scenarios, since the attacker assumes not only having access to the training data, but also depends heavily on prior knowledge of the feature space and the clustering model itself.

\subsection{Attacks using gradient optimization in NN}
Muñoz-González et al. \cite{Munoz_1} use a back-gradient optimization to perform poisoning attacks on DL models (see Figure \ref{DNN}). The gradient is calculated using automatic differentiation. Also the learning process is reversed in order to lessen the complexity of the attack. Such poisoning attacks display a wider capability of attack, able to be employed for multi-class problems rather than only binary classification. In addition, the poisoning examples entailed in the training phase do offer an adequate generalization over diverse learning models. This kind of approach has proven to be effective when dealing with handwritten recognition problems, spam filtering, and malware detection. Nonetheless, by performing back-gradient optimization to generate one poisoning data each time, the demand of even higher computation power increases.

\Figure[!h][width=0.4\textwidth]{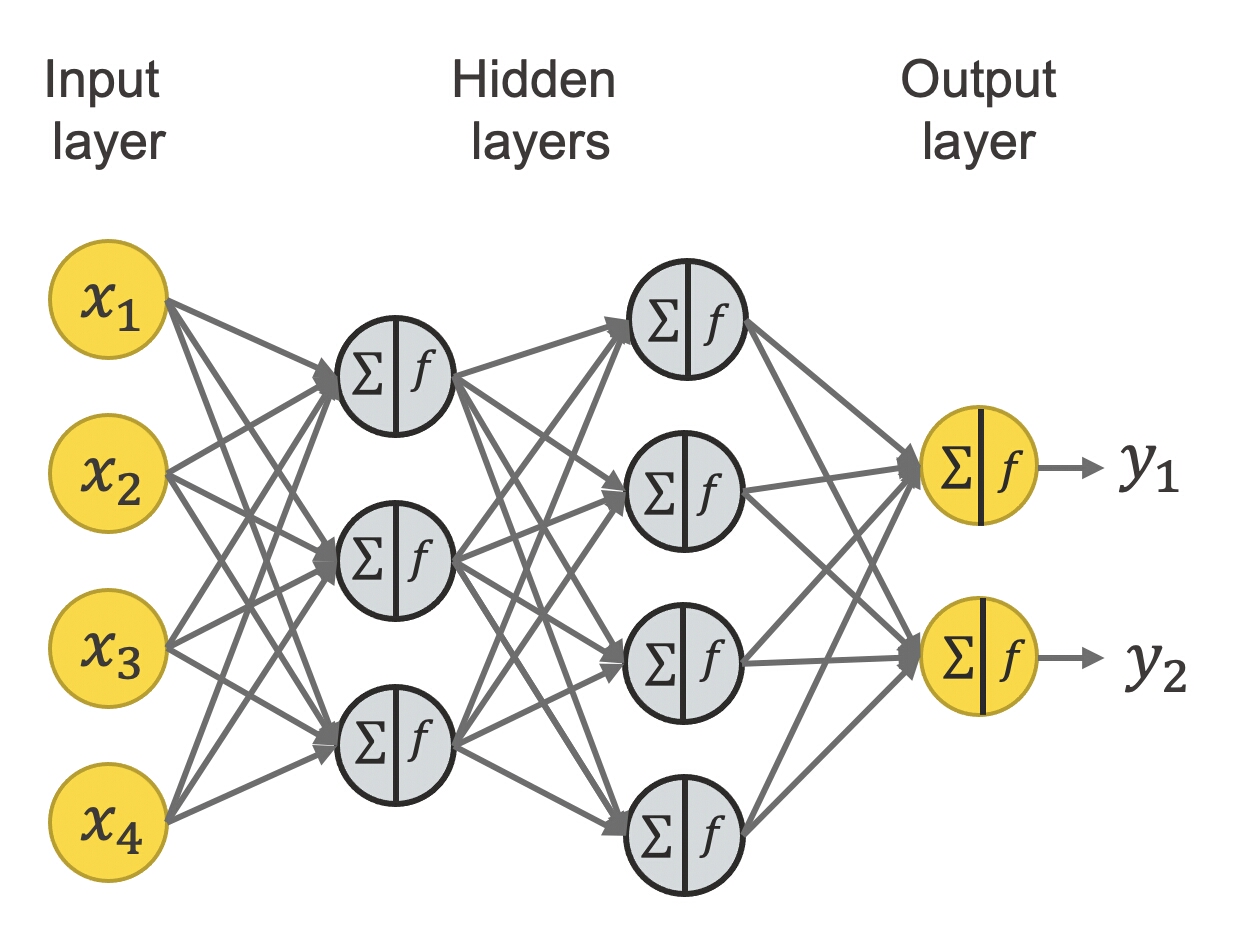}{Framework on a deep neural network \cite{A11}. \label{DNN}}

Yang et al. \cite{C_Yang} addresses the possibility of generate poisoning data targeting NN by harnessing the power of traditional gradient-based methods, or direct gradient method, via leveraging the weights of the target model. This work proposes two poisoning methods, the first one involving a direct gradient method and the second an autoencoder/generator of poisoning data. The generator approach with an autoencoder acting as a generator being updated via a reward function resulting from the loss. Then the model being targeted for such an attack becomes the discriminator. Once the discriminator/target model receives the poisoning data to compute the loss with respect to the normal data. 

\ 

The auto-encoder (or generator) based method exceeds in accelerating the generation rate of poisonous data by up to 239 times faster than relying on the direct gradient method. However, direct gradient method achieves 91.1\% of accuracy degradation while the generative approach is about 83.4\%, representing just a minimal decline in this metric. In addition, the generative method is proven to be superior in matters that regard poisoning larger NN models and bigger datasets. Such a difference is noticeable after conducting tests not only on MNIST dataset \cite{MN}; but also using CIFAR- 10 dataset \cite{CIFAR}, in which both poisoning schemes obtained similar accuracy degradation.


\subsection{Attacks using GAN (Generative adversarial networks)}
In a very parallel fashion to \cite{C_Yang}, Muñoz-González et al. \cite{Munoz_2} devices a optimal poisoning named pGAN, consisting of a generator that crafts the poisoning data; capable of both maximizing the error of the target classifier model, to degrade its performance, and achieving undetectability against any potential defense mechanism. The later features allows the proposed mechanism to be tested with ML classifier and NN, modeling the attacking strategy with varios levels of aggressiveness, also model the attack based on different detectability constraints \cite{Paudice_2} against some potential mitigation actions from the target model, testing then the robustness of the algorithm in question. The generator produces poisoning data based on the maximization of both the discriminator’s loss and the classifier’s loss on the poisoning data points. The labor of the discriminator entails distinguishing honest data and the generated poisoning data. The classifier purpose is minimizing a portion of the loss function containing a minor portion of poisoning data points throughout the training phase, identifying regions of the data distribution with prone to vulnerabilities and yet more challenging to detect. 

\ 

Based on the interaction of these three elements: generator, discriminator and classifier, a trade-off between detectability and attack effectiveness is achieved and measured by the computation of a hyperparameter alpha, indicating the probability of the crafted data to evade detection, the lower its value, the more accuracy drop but the more chances of the attack being detected, thereby the key elements consists on controlling the value of this hyperparameter. Experimental tests conducted on MNIST \cite{MN} and Fashion-MNIST \cite{FMN} show the effectiveness of the proposed poisoning technique assuming 20\% of poisoning samples. In contrast, accuracy degradation using pGAN is less comparing it with plain label-flipping techniques. Nonetheless label-flipping ignores any detectability constraints, therefore label flipping can only outperform pGAN at the non-existing presence of a defense mechanism. 

\ 

Chen et al. \cite{J_Chen_2} proposes DeepPoison as stealthy feature-based data poisoning attack, capable of generating poisoned training samples indistinguishable from the honest samplies for human visual inspection, then making the poisoned samples mush less identifiable throughout the training process. Also the scheme proposed displays high resistance against other defense methods, since many existing defenses account for attack success rates deployed with patch-based poisoning samples. The scheme is based on a GAN composed of one generator of poisoning data and two discriminators; one of the discriminators sets the ratio of the poisoning perturbations, while the second emulates the target model to test the effects of poisoning. After certain evaluations on public available datasets (Labeled Faces in the Wild\footnote[1]{http://vis-www.cs.umass.edu/lfw/} and CASIA\footnote[2]{http://www.cbsr.ia.ac.cn/english/IrisDatabase.asp}), DeepPoison shows to attain a maximum attack success rate of 91.74\% during testing phase, this by deploying only 7\% of poisoned samples. The performance is also analized when performed against two anomaly detection defense mechanisms: autodecoder defense \cite{M_Du} and cluster detection \cite{J_Chen_1} showing no steep drop in the ASR (maximum around 10\% drop).

\subsection{Feature-based poisoning attacks}

Feature-based poisoning attacks could accomplish positive aspects in terms of privacy preservation, since the poisoned training samples remain indistinguishable from the honest samples from the human visual perspective as seen in \cite{J_Chen_2,J_Shen}. This property surges as another topic of interest in devising a poisoning attack that can serve against abusive data collection which often represents a risks of user privacy violations. A potential application of such approach is mentioned in TensorClog \cite{J_Shen}, an attack could be intentionally directed to the media content of an smartphone gallery app just before sharing this data via social media, all of this without affecting the human perception of the media content, this mechanism is shown in Figure \ref{Tensor_Social}.

\Figure[!h][width=0.35\textwidth]{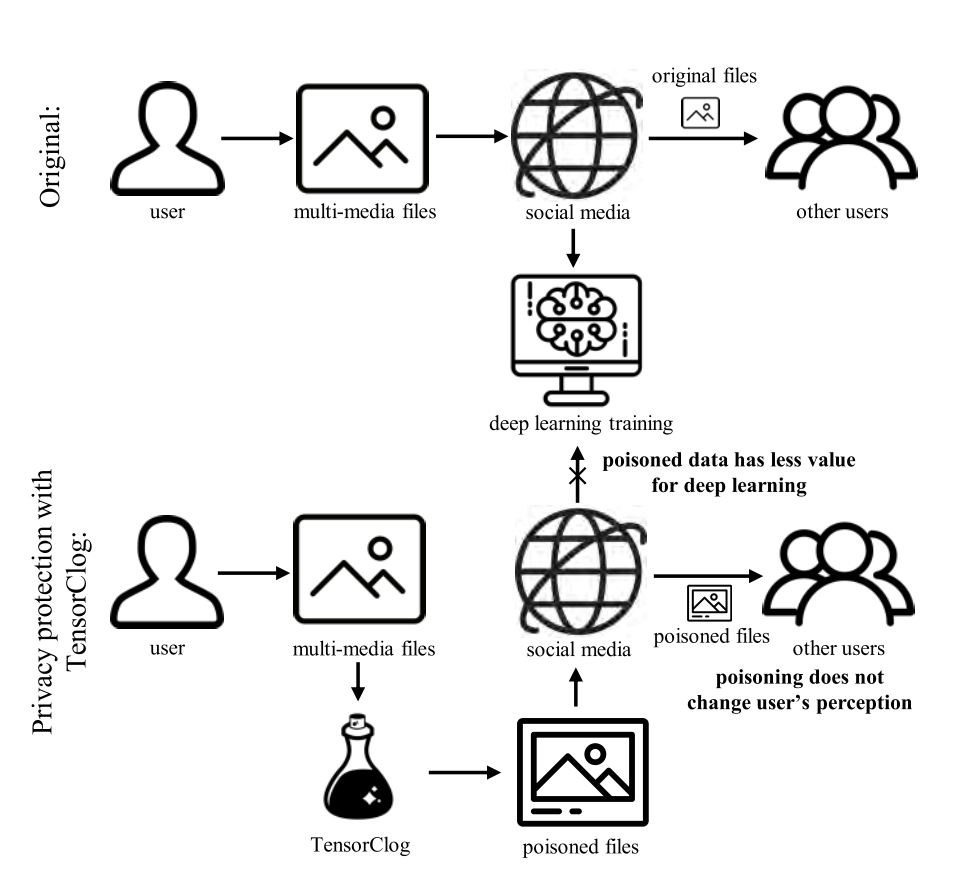}{Privacy protection scheme with TensorClog \cite{J_Shen}. \label{Tensor_Social}}


TensorClog poisoning attack \cite{J_Shen} attains the degradation of the overall accuracy of the target machine learning model by increasing the training loss by 300\% and test error by 272\% while preserving a high human visual similarity of SSIM $= 0.9905$, index that estimates quantitatively the change of human visual perception when assessing an existing similarity between two images. Being the previous results the product of an experiment conducted in a real life scenario over the CIFAR-10 dataset \cite{CIFAR}. The TensorClog attack technique is the result of clogging the back-propagation for gradient tensors, minimizing the gradient norm, throughout the training phase. Such a minimization of the partial derivative associated to the lost function weights practically leads to a deliberately caused gradient vanishing, jeopardizing the training process obtaining a larger converged loss. Furthermore, it succeeds in regularizing the added perturbation in the dataset without compromising the same; avoiding any possible human imperceptible perturbations by regularizing the distance between the poisoned and the clean samples. The effectiveness of TensorClog attack depends strongly on the availability of some of the target machine learning model information such as: Input-output pairs, model architecture, pre-trained weights, initialize function for the trainable layer, initialized value of the trainable layer. Having full access to all of the 5 key elements mentioned above indicates an attack on a white-box model/assumption; otherwise described as black box, wherein the effectiveness of the poisoning attack declines drastically.

\subsection{Attacks on Crowd-sensing systems}
Crowd-sensing systems are vulnerable to data poisoning due to the dearth of control over the worker’s identities. Each poisoning strategy is based on creating interference with the collected data via fake data injection. Li et al. \cite{M_Li} addresses vulnerabilities issues related to crowdsensing systems, such as TruthFinder’s framework, by developing a poisonous attack strategy that considers access to local information (obtained through the attacker’s sensing devices and the results announced by TruthFinder) rather that global information (overall distribution of data including honest and poisoner workers).
Modeling the proposed scheme as a partially observable data poisoning attack based on deep reinforcement learning, allowing the poisoner workers to tamper with TruthFinder and remain hidden within the network, the optimization strategy enables the poisoner workers to learn from past attack attempts and progressively improve. A series of conducted experiments showcase the performance of the proposed attack strategy called refered to as ‘DeepPois’ and tested alongside another two baseline models, the metric to compare is in terms of the rewards yielded by each method, the cumulative reward as per every episode in DeepPois reaches superior values than the others. Moreover, the cumulative reward increases as the number of episodes does since DeepPois considers historical attack experiences in the learning process.

\subsection{Attacks on data aggregation models}
Zhao et al. \cite{P_Zhao_1} analyzes poisoning attacks occurring on data aggregation, commonly refer as garbage in, garbage out; as representative example, an attacker could transmit to the aggregator a set of poisoned locations. This work focuses on the inputs of data aggregation and by so this work proposes a novel attack framework capable of deploying poisoning attacks on location data aggregation (PALDA); in addition, PALDA accomplishes to disguise the behavior of each launched attack. The entailed process requires to define the poisoning attack as a min-max optimization problem to solve via an iterative algorithm, which has been proven theoretically to achieve convergence. The poison strategy entails on tampering  the aggregated results at the output of the aggregation model obtained by the aggregator; this in form of the an error, obtained when comparing the existing similarity in terms of mobility between honest with the poisoned locations as the aggregated results. Then PALDA minimizes the aggregation parameters of the model and maximizes both the error of the aggregated results at the output of the aggregation. Also, the attack strategy of this approach is presumed to be extended to other settings, representing a suitable option to linear decomposable aggregation models that are executed by an aggregator. The effectiveness of the proposed poisoning attack is simulated onto six different GPS mobility datasets. Three of them provided by Microsoft Research Asia, featuring the cities of Beijing, Hongkong and Aomen; also the loc-Gwalla and loc-Brightkite datasets (location-based social networks) and the Athens truck dataset. A benchmark evaluation is conducted by comparing the performance of PALDA alongside with the poisoning attack named Synthesizing \cite{Bindschaedler} and Baseline, the later consisting on poisoned locations randomly generated being fed to the aggregator. During the evaluation it is observed a drastic increase in the MSE when running PALDA scheme over the other two poisonous schemes. Nonetheless, the entire evaluation is conducted assuming poisonous rate of less than 20\%.

\ 

The false positive and false negative rates are experimentally tested for the three schemes, observing a superior numbers with PALDA. Attributing then to PALDA the capability of disguising the behavior of each launched attack, feature not present in the other two schemes. Making then even harder for a defense system to detect such threads, being honest users more likely to be considered poisoners; which is mainly due to the algorithm optimizing approach. Developing in the nearest future a defense system against PALDA attacks cannot be successful by pondering the existing physical proximity among honest users in order to detect the attackers, because PALDA succeeds in generating users with a higher probability of accomplishing physical proximity to honest users. Then a better suggestion for a poisoning defense could be presented in the form of a method that entails simultaneously both physical proximity and social relations as a network, assuming by then no existing social interaction processed as side information between attackers and honest users.

\subsection{Miscellaneous attacks} 

\textbf{Attacks on principal component analysis:} Rubinstein et al. \cite{Rubinstein} proposed a series of poisoning attacks on Principal Component Analysis (PCA), this by inserting a minimum portion of poisoned data; and as a result, the performance of the detector diminishes drastically. However, the success of this approach relies on binary classification algorithms, leaving aside any possibility of being deployed on generic models nor learning algorithms.

\ 

\textbf{Attacks against an specific defense:} Koh et al. \cite{Koh} proposes a DP attack against a defense mechanism considered in \cite{Steinhardt}. This scheme considers prior knowledge of the sanitization defense as the main assumption that allows it to tamper the accuracy of the ML model. This approach is successful at evading the defenses since the poisoned samples are placed in close proximity to the honest data; therefore result, they cannot be considered outliers. However, this attack scheme is heavily dependent on the attacker knowing both the training and the test dataset.

\Figure[t!][width=0.95\textwidth]{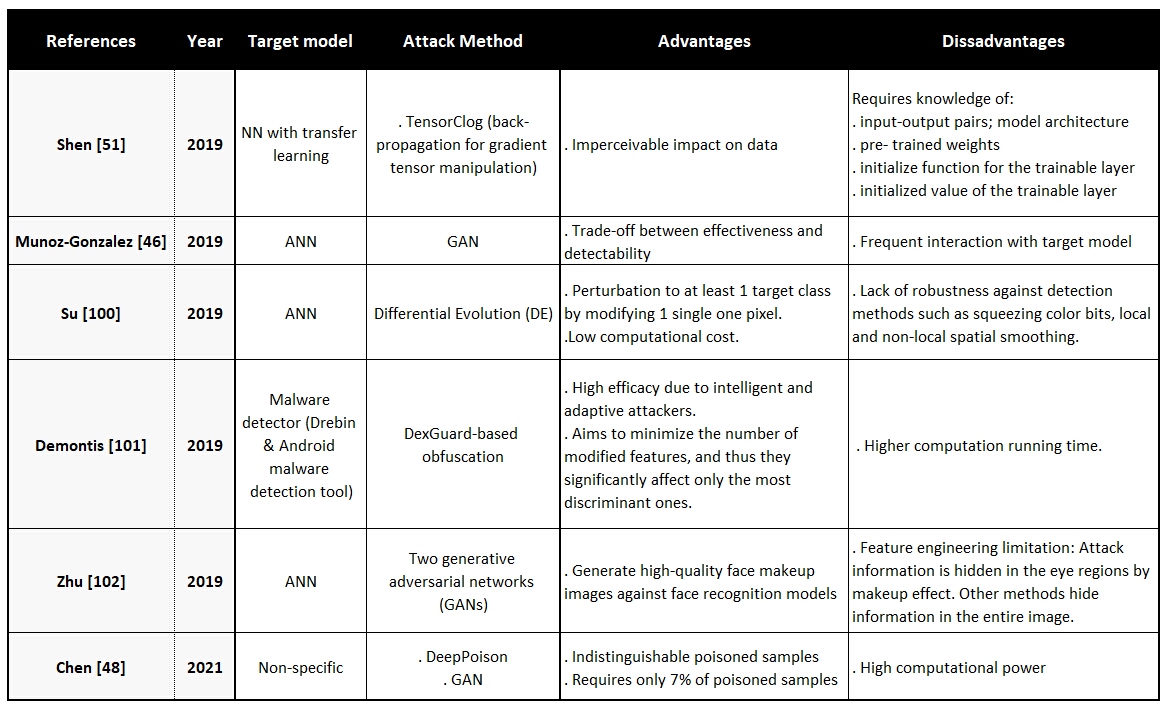}{The poisoning attacks (2019-2021).\label{attack_table}}


\section{Defense Mechanisms}
In this section, several defense techniques for ML are reviewed. Covering various countermeasures and different approaches published over the most recent years with the aim of diminish the damage/possibility of a poison attack. These can be divided in two kinds, protection for the data and protection for the model, this last category has been segmented based on whether the target model is a NN model or a non-NN model. 

\subsection{Defenses in Collaborative Learning}
With a data poisoning attack an attacker aims to tamper the integrity of a machine learning model or cybersecurity system by maliciously modifying its behavior by modifying any of the data examples \cite{Gu} or by simply changing the labels of the training data \cite{Bhagoji}. These kinds of defense mechanisms aim to mitigate the DP based on collected data \cite{Baracaldo}. Collaborative Learning is especially vulnerable to poisoning attacks due to the fact that the servers, while generating the necessary updates, lacks of any ability to look into the process. Zhao et al. \cite{L_Zhao} presents a novel defense approach capable of detecting anomalous updates, the scope of this application covers iid (independent and identically distributed) and non-iid settings. After performing an evaluation on MNIST \cite{MN} and KDDCup datasets \cite{KDDC}, the proposed solution is proven to be robust enough against two distinctive label-flipping poisoning attacks. Upon realizing client/side cross validation, the detection task is assigned to the clients in order to evaluate the performance resulting from each update, which at the same time is evaluated over other’s client’s local data. When performing aggregation, according to the evaluation results, the server is able to modify the weights associated to the updates. The limitation of this scheme is associated to a considerable computational costs for both performing the testing on the updates on the client side and also transmitting the sub-models to the clients.

\ 

\subsection{Defenses in Federated Learning}
Federated Learning (FL) in recent years has become relevant in privacy-preserving applications. This is possible since the data gathered from each device or worker is kept locally stored in each device \cite{H_Xu}, then enabling the training process of a sub-machine learning model individually. As a second step, only the resulting gradients obtained after training are exchanged to a centralized server instead of the raw data, then the centralized server performs the entire training lifecycle by multiple iterations until attaining a desirable accuracy. Due to the nature of FL, malicious users could perform a label-flipping attack \cite{B_Biggio_3} by deliverately inserting crafted gradients leading to classification errors during the test phase. In the past it is been proven that a single poisoner can undermine the whole training process and as a result the integrity of the model. Therefore, a robust FL model needs to regard on concerns related not only to data privacy, but also rely on a certain degree of resilience against poisoning attacks and data manipulations. Liu et al. \cite{X_Liu} addresses the privacy/defense related issue in FL models by showcasing a novel framework called privacy-enhanced FL (PEFL). PELF grants the central server the ability to detect malicious gradients and block poisoner workers. By comparing the malicious gradients, submitted by the poisoner workers, as a set of parameters to the same ones belonging to the honest workers; the difference between malign and benign gradient vectors can be evaluated by calculating the Pearson correlation coefficient \cite{Benesty}. Abnormality behavior is related to a lower correlation coefficient, then the action of the defense mechanism consists on simply setting the weights of the malign model to zero. PEFL claims superiority among other similar systems such as Trimmed Mean \cite{H_Xu}, Krum \cite{Blanchard} and Bulyan \cite{Guerraoui}. Since the proposed scheme does not assume to have any knowledge of the total number of poisoners, posing then a more appropriate defense more suitable for real-case scenarios. Furthermore, PEFL poses a higher resilience against accuracy drops compared to Bulyan and Trimmed Mean due to weight adjustment performed on each gradient which guarantes trustworthiness within the remaining parameters. Observing in the end a maximum attack success rate of 0.04, evidence of the robustness of the model against label-flipping. 

\subsection{Miscellaneous defense mechanisms} 

\textbf{SVM Resistance Enhancement} \cite{Weerasinghe} is targeted to avoide label-flipping attacks, being SVM particularly vulnerable against this kind of attacks, causing total misclassification due to the computation of erroneous decision boundaries. Thinking ahead about the effects of suspicious data points within the SVM decision boundary, the proposed approach considers a weighted SVM accompanied by KLID (K-LID-SVM). This work introduces K-LID, a new approximation of Local Intrinsic Dimensionality (LID), metric associated to the outliners in data samples. K-LID computation relies on the kernel distance involved in the LID calculation, allowing LID to be computed in high dimensional transformed spaces. Obtaining by such means the LID values and discovering as a result three specific label dependent variations of K-LID capable of counter the effects of label-flipping. K-LID-SVM attains higher overall stability against five different label-flipping attack variants: Adversarial Label Flip Attack (alfa) attack, ALFA based on Hyperplane Tilting (alfa-tilt), Farfirst, Nearestl and Random label flipping; using five different real-world datasets for a benchmark test: Acoustic, Ijcnn1, Seismic and Splice and MNIST. The defense system attains a drop of 10\% on average in misclassification error rates, this method can distinguish poison samples from honest samples and then suppress the poisoning effect. Therefore, it succeeds in decreasing the potential magnitude of the attack significantly and demonstrating a superior performance than traditional LID-SVM.

\

\textbf{Bagging classifier} \cite{B_Biggio_4} is an alternative to detect poisoning samples, representing this method an ensemble method that accomplishes the negative impact of the outliers influence over the training data. Then Biggio addresses the existing similarities between DP attacks and outlier detection problems, assuming a reduced number of outliers with shifted distribution behavior. The training of this ensemble method requires different training samples and considers multiple classifiers as well. Thus the combination of the all the predictions obtained from the multiple classifiers can be harnessed in order to mitigate the strength of the poisoning data/outliers. This defense approach is evaluated over two scenarios, one including a web-based IDS and a spam filter. However, a considerable computational power is demanded to deploy the proposed defense system. 

\ 

\textbf{Kernel-based SVM} \cite{H_Xiao_1} has been proposed to combat DP attacks that entail mislabeling actions. This approach showcases signs of improved robustness when diminishing the SVM slackness penalty, present on margin violations, enabling a higher number of samples to take part in the estimation of the SVM’s weight vector. Albeit, the most significant contribution in this work is in attaining the substitution of the existing dual SVM objective function with an dual function that is rather based on probability of accounting for mislabeling rather than on the training labels. 

\ 

\textbf{ANTIDOTE} \cite{Rubinstein} is a defense scheme to counteract the actions of poisoning attacks treating anomaly detector. Based on a statistical approach, ANTIDOTE is able to weaken the effects of the poisoning attack by discarding the outliers present in the training data. 

\ 

\textbf{KUAFUDET} \cite{S_Chen} is a defense strategy to combat DP in malware detection. This defense system employs a self-adaptive learning framework to detect and avoid suspicious results falling in the category of false negative to take part of the training process.

\

\textbf{De-Pois} \cite{J_Chen_3} is an attack-agnostic defense system against poisoning attacks named De-Pois. The defense strategy depicted in this work is not to attack-specific, i.e. it is not designed to combat one type of attack in specific, but assert in attaining more generality over its deployment than other defense techniques. Regardless previous knowledge on the type of machine learning model being targeted, nor the type of assumptions taken by the attacker; De-Pois scheme can discriminate between the poisoned data samples from the honest samples is achieved based on the results of a prediction computed to compare both the target and the mimic models.

\Figure[!h][width=0.9\textwidth]{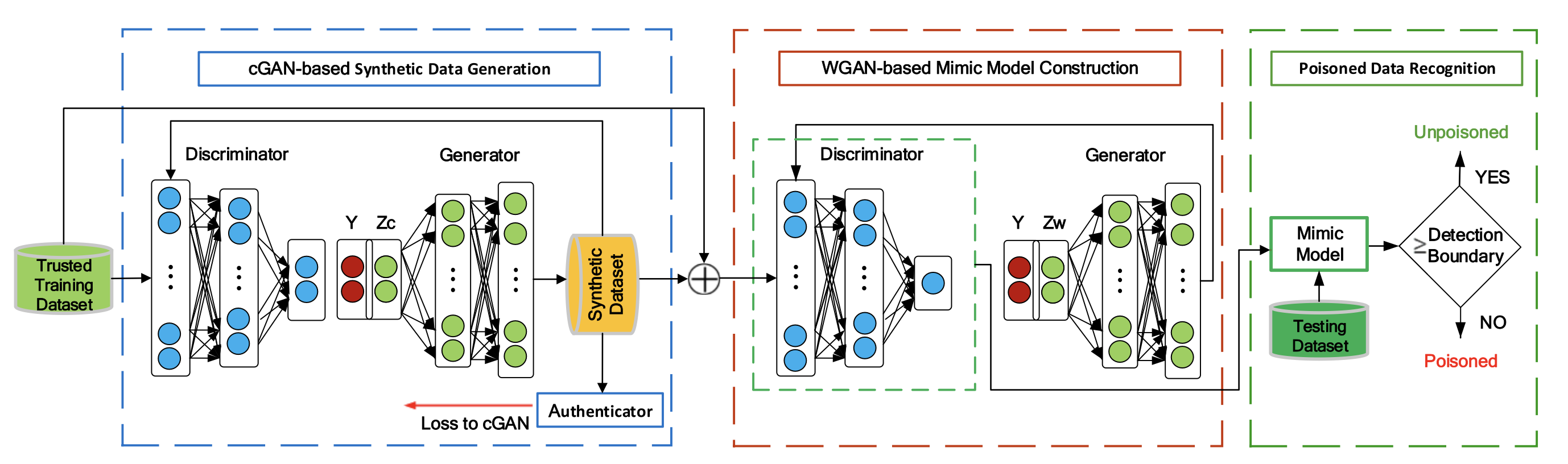}{Configuration of De-Pois framework: cGAN-based for synthetic data generation,  WGAN-GP for mimic model construction and poisoned data recognition \cite{J_Chen_3}. \label{De-Pois}}

Firstly, De-Pois generates sufficient synthetic training data that resembles a similar distribution to the honest data samples by employing a cGAN (conditional GAN) to capture the distribution of the clean data \cite{Mirza} and then it is used for data generation and the inclusion of a discriminator to monitor the data augmentation process. Afterwards, a conditional version of WGAN-GP (Wasserstein GAN gradient penalty) \cite{Gulrajani} is set to learn the distribution present in the predictions related to the augmented training data, the mimic model is obtained by extracting the discriminator part from WGAN-GP as shown in Figure \ref{De-Pois}. As a result, the mimic model attains a similar prediction performance to the target model. Finally, by employing a detection boundary the poisoned samples can be set apart from the honest samples. Then gauging the output from the mimic model with the obtained detection boundary a sample can be regarded as either poisoned or honest. 
The effectiveness of De-Pois is tested against various poisoning attacks including CD \cite{Steinhardt}, TRIM \cite{Jagielski}, DUTI \cite{X_Zhang}, Sever \cite{Diakonikolas} and Deep-kNN \cite{Peri} by applying commonly known image sets including MNIST and CIFAR-10. De-Pois succeeds in detecting poisonous data when facing every distinctive attack scheme; performing better than most attack-specific defense systems, obtaining an average value of over $0.9$ for both accuracy and F1-score. 

\ 

\textbf{k-Nearest-Neighbours defense scheme} \cite{Paudice_2} is designed to detect malicious data and counteract the effects of the same, being this defense referred as Label sanitization (LS). Label sanitization (LS) bases its defense on the decision boundary of SVM, observing the remoteness of the poisoned samples, commending these samples to be re-labelled. Steinhardt et al. \cite{Steinhardt} proposes a nearest-neighbor-based mechanism to detect outliers and SVM optimization right afterwards, getting as a result a domain-dependent upper bound associated to the estimated highest drop in accuracy due to a DP attack. A special assumption is made for this scenario, declaring the removal of nonattack outliers inconsequential to the performance of the target model. 

\

\textbf{Active Learning Involvement:} Active learning systems entails the intervention of a human expert commonly referred to as an “oracle”. As a result, the number of labeled samples is increased by preforming a selection of the samples from unlabeled batches, supposing enhanced results by just performing libeling, most of the times labeling the samples associated to higher uncertainty values (closer to the decision boundary). The main concept is to allow the model to be trained again with these set of new labeled samples. Some researchers suggest suggests a selection scheme that considers a mixed sample scenario; in which, during the labeling step, the selection process is performed in a random manner \cite{D_Miller}. The target model is then prone to generalize better against DP by the introduction of a certain level of randomness. Such a strategy promises a significant reduction in the frequency of the attacker’s labeling attempts, diminishing the resulting accuracy degradation. Human-in-the-Loop is targeted to improve the accurate rate detection of DP attacks can be addressed by the inclusion of a human expert in the loop \cite{B_Miller}. Such approach is useful for situations where the detected attack requires further characterization in order to propose a set of response actions. Then, rather than simply detecting anomalies, the aim is to categorize these; for instance, impeding the access to the same attacker. In the long-run, a Human-in-the-Loop defense enables the integration of active learning towards the inclusion of an automated classifier capable of mimicking and eventually replace the human expert.

\Figure[t!][width=.95\textwidth]{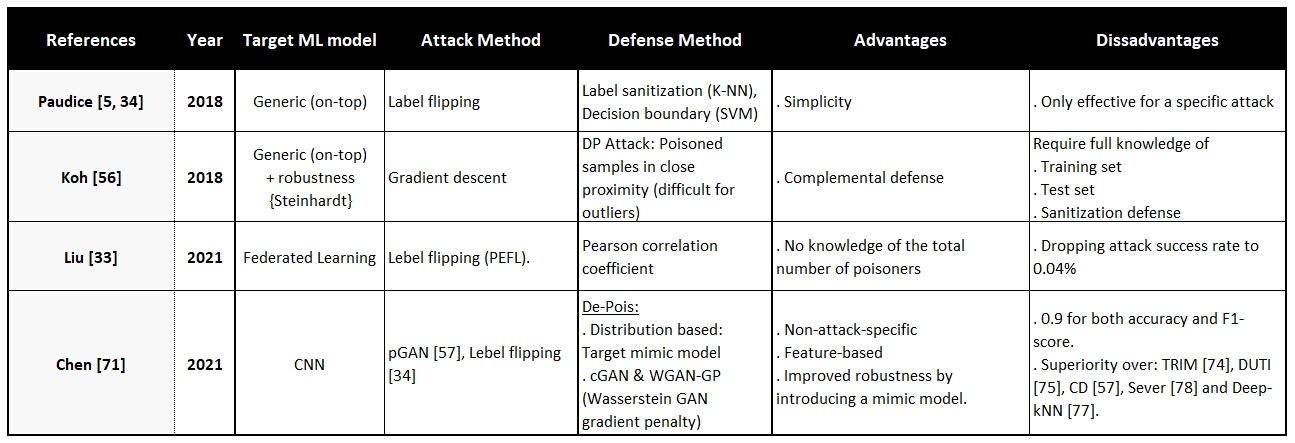} {The defense techniques (2018-2021).\label{defense_table}}

\section{briefs of our survey studies}
This section is mainly summarizing the contents what the current studies in the survey have mentioned. The selected discussion points and/or further research considerations are covered in this section. 

\subsection{Discussion Archives}
It is important to highlight the attacker’s capability in other scenarios and the assumptions imposed by the attacker, being sometimes to optimistic while in other scenarios represent more realistic conditions \cite{J_Chen_3}. In the literature, there are many examples related to assumptions in regards to the capability of the attacker. For instance, TRIM \cite{Jagielski} assumes the ratio of the poisoning examples declared by the attacker as known. Deep-kNN \cite{Peri} assumes access to ground-truth labels allowing the system to compare each sample’s k neighbors with the class labels. CD \cite{Steinhardt} considers the influence of the outliers as inconsequential to the drop in performance of the target model. Based on this assumption, an approximation of the upper bounds about the loss is tested against poisoning attacks with non-convex settings. It is important to make a clear distinction on the properties of the crafted poisoning data, poisonous data obtained by performing label-flipping is one. Albeit, the scheme promises high accuracy degradation, it is far from representing the most effective option for an attacker. This is because label-flipping is considered among the most basic poisoning techniques and most of the existing defense mechanism can detect these ones as outliers and reject them with relatively ease. 

\ 

On the other hand, other poisoning attacks are feature-based, focusing solely on the sample’s feature representation and by so relies on the distribution of the training data to generate poisoning samples. This approach is more efficient since it does not rely on attacking capability, then the number of samples necessary to cause the target model to drop in accuracy is minimum. A clear example of this is shown in DeepPoison \cite{J_Chen_2} where only 7\% of poisoned samples were required to cause an devastating 91\% drop in accuracy, demostrating superior robustness when compared to other poisoning schemes such as: BadNets \cite{Gu}, Poison Frog \cite{Shafahi}, Invisible Poisoning attack \cite{J_Chen_1}, Fault Sneaking attack \cite{P_Zhao_2} and various backdoor attacks \cite{X_Chen, A_Saha, B_Wang_2}.

\ 

Powerful poisonous attacks data generation based on GAN such as \cite{Munoz_2} and \cite{J_Chen_2}, both present an attractive option that offers high accuracy degradation with minimum attacking capability, in some cases promising a much reduced training time compared to other schemes, such as the direct gradient method. Albeit, employing a GAN implies a higher computational overhead, direct gradient method attains a higher accuracy degradation than the GAN approach, as seen in \cite{C_Yang} GAN-based poisoning attacks can deliver equal to results when dealing with a significantly bigger training set. Most of the defenses already explored in this work propose an end-to-end solution to counteract DP attacks based on a detection strategy, once the poisoning samples are identified they are rejected from the training set, thus nullifying the attack. Another approaches succeed in making the models more robust against the effects of outliers, considered as poisoning samples \cite{Rubinstein, B_Biggio_4}. Nonetheless, significant computational overhead is demanded continuously, also the three methods limit themselves to binary classification tasks.

\subsection{Suggested Research}
Recent ML models can no longer be seen as black box systems to be assessed solely on their results, but a deep understanding of the model is necessary to identify security flaws that could lead to critical undesirable outcomes if not properly addressed. Any breakthrough in the field of machine learning will always implicitly allow the introduction of new vulnerabilities. Such vulnerabilities will always pose an open window of opportunity for adversarial entities to exploit, leading as a consequence to an opportunity to develop new defenses counterbalance the outcome in the same matter as it occurred with the invention of the internet and the introduction of cybersecurity systems.

\ 

Ensemble methods represent another approach of potential interest capable of detecting  poisoning attacks over different ML model architectures (ex. smart devices on crowd sensing systems) where a partition of the original dataset is given to every ML model to train with. A potential solution to this problem would imply sending the local data from one smart device to other devices that conform a dynamic ensemble, this in order to perform predictions on the data received and label them as either benign or malign before submitting it to a query device. There are other metrics that imply a more behavioral approach and can play an important role in the prediction process and the selection process of the devices as part of the ensemble to indicate the existence of a poisoning attack threatening the model, a representative example would be a trust/confidence score. 

\ 

The introduction of a unified method capable of performing an exhaustive security evaluation over the model’s robustness has not been formally attained. It is believed that, if such method were established, the effects and implications resulting from the interaction of both attack and defense strategies would be better assessed in throughout every experiment \cite{M_Xue}.
Such comprehensive approach could be extended onto privacy related evaluations, focusing specially on the training set and the hyperparameters associated to the target model. Albeit, \cite{B_Biggio_1} proposes a method that could address partially this need of comprehensive method, there is an wide and growing area of opportunity in this matter.

\section{Survey Analytics and Insights}
This section is the report of the survey results which includes the general statistics and the major insights of the data poisoning attacks and their defense mechanisms. The qualitative survey analysis for broad ranges of the studies is one of major contributions of this survey.

\subsection {Survey Statistics}
More than 80 research papers (out of 106 references) have been carefully reviewed for this survey study. According to our survey, the research contributions which are related with the data poisoning and its defense mechanisms are appeared in the mid 2000 and have been widely studied in the late 2010 (see Figure \ref{Stat01}). The related surveys \cite{A01, Miller_Survey, Liu_Ximeng, He, M_Xue} in theses research topics are also appeared after 2018 (orange color in Figure \ref{Stat01}).

\Figure[h][width=0.45\textwidth]{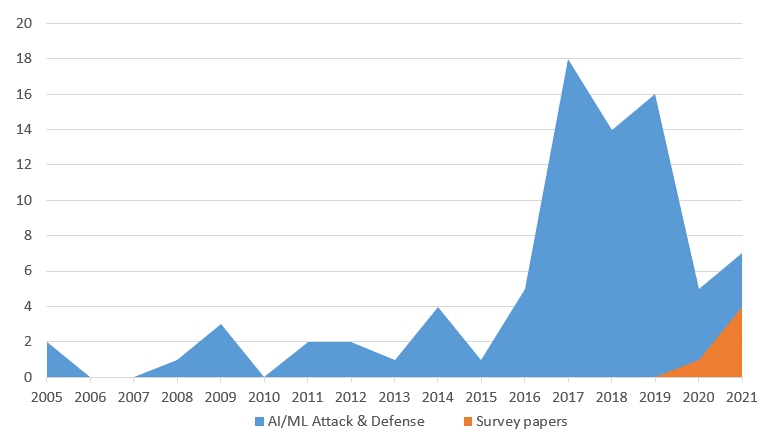}{Number of publications on the survey. \label{Stat01}}

Around $65 \%$ (i.e., $54\% + 11\%$) of the research is about the data poisoning attack methods and around $46 \%$  (i.e., $35\% + 11\%$) of the research is about the defense mechanisms. It is noted that some studies covers both attack and defense together (A/D) which is around $11 \%$ (see Figure \ref{Stat02}). One of interesting insight is that more than $65 \%$ of papers are registered on \textit{ArXiv.org} which is an open-access repository of research articles in the fields of mathematics, computers, sciences and engineering \cite{AXV}.

\Figure[h][width=0.45\textwidth]{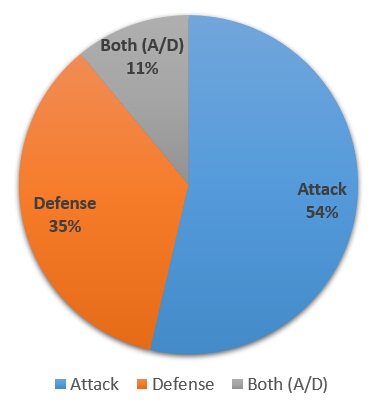}{Category of the survey. \label{Stat02}}

\subsection {Defense mechanisms for testing nodes}
According to our analysis, there are no data poisoning defense mechanisms (see Figure \ref{defense_table}) in the testing phase and this result is aligned with the previous research \cite{Liu_Ximeng, M_Xue}. Most of defense methods are targeted to protect the integrity of testing nodes (i.e., single or multiple trained machine learning models). A trusted execution environment (TEE) is a secure execution environment (mostly connected hardware components) which guarantees code and data loaded inside to be protected with respect to confidentiality and integrity \cite{Liu_Ximeng}. Although dedicated TEEs for AI systems have been widely studied \cite{Q_Xia, B_Wang_1, Trame, Juuti}, the mechanisms for constructing secured environments are broader than just AI dedicated ones. Non-AI system dedicated defense mechanisms could be applied to protect trained machine learning models as the TEE in the testing phase \cite{Z_Chen}. Even error detection method for a ML system could be also adapted for developing a TEE as a defense mechanism \cite{Chakarov}. The blockchain based defense mechanisms have been proposed for guaranteeing the integrity within connected nodes (i.e., network system components) \cite{B01, B02, B03}. These techniques could be also applied into AI systems as the defense mechanisms. It is noted that any security mechanisms for improving integrity (and confidentiality) of systems could be considered as AI system dedicated TEEs. In the other hand, the malware detection for AI models in the testing phase is also another typical defense mechanism to protect testing nodes \cite{R07, R01}. The TEE and the malware detection are two major categories for the defense mechanism in the testing phase \cite{Liu_Ximeng}.

\subsection {Attack Cost for Machine Learning Systems}
Training models equipped with poisoning detection techniques can either increase the resilience of a classifier against the insertion of training poisoning data. As a result, new ML models are expected to develop a certain degree of immunity. This will open a debate on the rising cost involved in the data poisoning process, which is predicted to be more costly in the years to come and could even discourage potential attack attempt to be crafted in the first place. The overall sophistication of the attacks has increased gradually over the recent years as a trend that is correlated to the resources available to the attacker; for example, employ more advance hardware such as GPUs. Then the resources of the attacker seen as a ‘budget’ is becoming a factor worth to be analyzed in the future, being predicted as an increasing trend in the years to come. Most successful attacks reviewed in this work consider the aim of the attacker to have an increased control over future classification over the model, based on this a common practice is to train a ML model that resembles and emulates the characteristics of the target model. This will not only serve to test the effect of the attack but also to approximate as close as possible to the separation surfaces of the classifier, having by then a low confidence point, as an initial point instead of starting at a random location without having any knowledge over the target model. Another example that associates the success of an attacker with its available resources is reflected on the capability of the attacker to limit itself to target linear models only or go beyond this scope.

\ 

\subsection {Neural networks and Adversarial Learning}
Neural network models showcase impressive performance across a wide number of applications, making it the algorithm of choice. Nonetheless NN poses a major flaw in terms of interpretability, which is a term used to describe the level of understanding over the decision process performed by a model on each prediction. Then interpretability poses a challenge difficult to overcome due to the nature of NN algorithms, since low interpretability impedes tracking predictive processes involving complicated logic and mathematical algorithms. High interpretability will then pose a more desirable scenario allowing the NN working mechanism to be effortlessly analyzed. Low interpretability leads to vulnerabilities related to privacy issues. Thereby it is foreseen that in the nearest future such a concern will be weighted even higher up to the point of being treated as an urgent matter. Nonetheless, any improvement in matters of interpretability can be harnessed not only by the defense, but by the attacker as well. Thereby, dept understanding of the model would facilitate an attacker to craft more effective adversarial examples and poisonous data. At some point, exploring new vulnerabilities could present itself as a more feasible option rather than detecting them prior to occur; thereby, giving the attacker the upper hand over the defender \cite{Veale}.


\section{Conclusion}
This paper offers a comprehensive survey on ML classifiers covering data poisonous attacks during the training phase, enlisting various types of attack schemes and countermeasures in forms of defense strategies. Different DP attacks and their associated defense strategies in various scenarios have been investigated and compared in terms of their performances, disclosing advantages and shortcomings for each approach. Stronger attack approaches have become a trend in the field of deep learning, showing remarkable potentials in recent years, that are capable of generating feature-based poisoning data that are found to be relatively more effective than other attack schemes. In addition, they target a wider variety of classifiers and other defenses as well, generally effective among diverse ML models. Moreover, the deep learning approach has shown concerning privacy related vulnerabilities which need to be addressed in the near future. Security related issues in machine learning still remain as an active research domain, which requires continued attentions from researchers in the area for the years to come. More complex security threats are predicted to emerge continuously which in turn requires developments of more advanced defense techniques to detect and counteract such threats.  Therefore, guarantying the robustness in any ML model against DP is foreseen to become a priority and become an industry standard.

\appendix

\EOD

\end{document}